\documentclass[journal]{IEEEtran}

\usepackage{cite}
\ifCLASSINFOpdf
  \usepackage[pdftex]{graphicx}
  \usepackage{epstopdf}
  \usepackage{color}
\fi
\usepackage[cmex10]{amsmath}
\usepackage{mdwmath, mdwtab}
\usepackage{amssymb}
\usepackage[normalem]{ulem}

\newcommand{\ds}{\displaystyle}
\newcommand{\bcenter}{\begin{center}}
\newcommand{\ecenter}{\end{center}}

\begin{document}
\title{Computing with Non-equilibrium Ratchets}
\author{Mehdi~Kabir*,
        Dincer~Unluer*, 
        Lijun~Li, 
        Avik~W.~Ghosh,
        and Mircea~R.~Stan
\thanks{All authors are with the Charles L. Brown Dept. of Electrical and Computer Engineering, University of Virginia, Charlottesville, VA 22904. This material is based on work supported by the Nanoelectronics Research Initiative (INDEX Center), ViNC, UVA-NanoSTAR, and UVA-FEST awards.
*These authors contributed equally to this work.}}
\maketitle

\begin{abstract}
Electronic ratchets transduce local spatial asymmetries into directed currents in the absence of a global drain bias, by rectifying temporal signals that reside far from thermal equilibrium. We show that the absence of a drain bias can provide distinct energy advantages for computation, specifically, reducing static dissipation in a logic circuit. Since the ratchet functions as a gate voltage-controlled current source, it also potentially reduces the dynamic dissipation associated with charging/discharging capacitors. In addition, the unique charging mechanism eliminates timing related constraints on logic inputs, in principle allowing for adiabatic charging. We calculate the ratchet currents in classical and quantum limits, and show how a sequence of ratchets can be cascaded to realize universal Boolean logic. 
\end{abstract}


\section{Introduction}
The exponential growth in computational processing has been driven by the scaling of CMOS technology to smaller and smaller scales. The largest impediment to sustained scaling is the large thermal budget arising from power dissipation in logic circuits \cite{zhirnov,itrs}. Biological systems perform switching operations quite efficiently, albeit at lower speeds and targeting very specific applications as opposed to universal logic. Nonetheless, the principles that many biological systems utilize are quite instructive - notable among them being the use of analog encoding of signals, and the employment of strongly non-equilibrium power sources. Brownian motors use non-equilibrium noise to create directed motion \cite{feynman,magnasco,astumian} to fuel the motion of motor proteins \cite{rice} and drive ion flow in artificial nanopores \cite{dekker}. The purpose of this paper is to explore the distinct  energy advantages of extending ratchet principles to solid-state electronics, specifically the shuttling of charges without a global drain bias under non-equilibrium conditions \cite{jung, borromeo, reguera}. 

We find two distinct advantages of an electronic ratchet related to its device level energetics. First, the absence of a drain bias needed to create global directionality reduces the static dissipation during device operation. In purely current-controlled logic (e.g. a Binary Decision Diagram or BDD), this is a distinct advantage. In a conventional CMOS incarnation however, charging an output capacitor converts this current source into an effective drain voltage. At this stage, the second advantage of a ratchet comes in - a voltage-controlled current source dissipates less energy when charging the capacitor, and is amenable to adiabatic charging. Conventional schemes for adiabatic charging require precise timing information for signal synchronization; since we wait till each capacitor charges up adequately, such a timing requirement is not demanded of our proposed ratchet. 

\begin{figure} [t]
\centering
\includegraphics[scale=0.3]{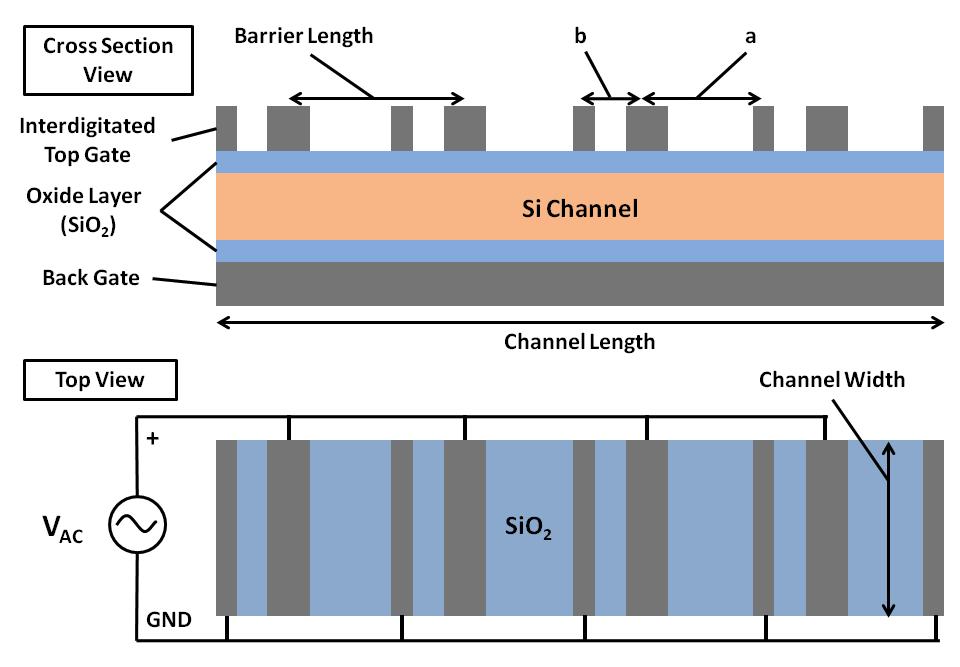}
\caption{Prototypical geometry of a ratchet consisting of an interdigitated top gate, which creates the asymmetric time varying potential, and a back gate, which determines the initial number of carriers in the channel by shifting the Fermi level. The current driven by the top gate is used to build up voltage, which can be used to back gate the next cascaded ratchet in the logic architecture.}
\label{fig:fig1}
\end{figure}

\begin{figure}[t]
\centering
\includegraphics[scale=0.25]{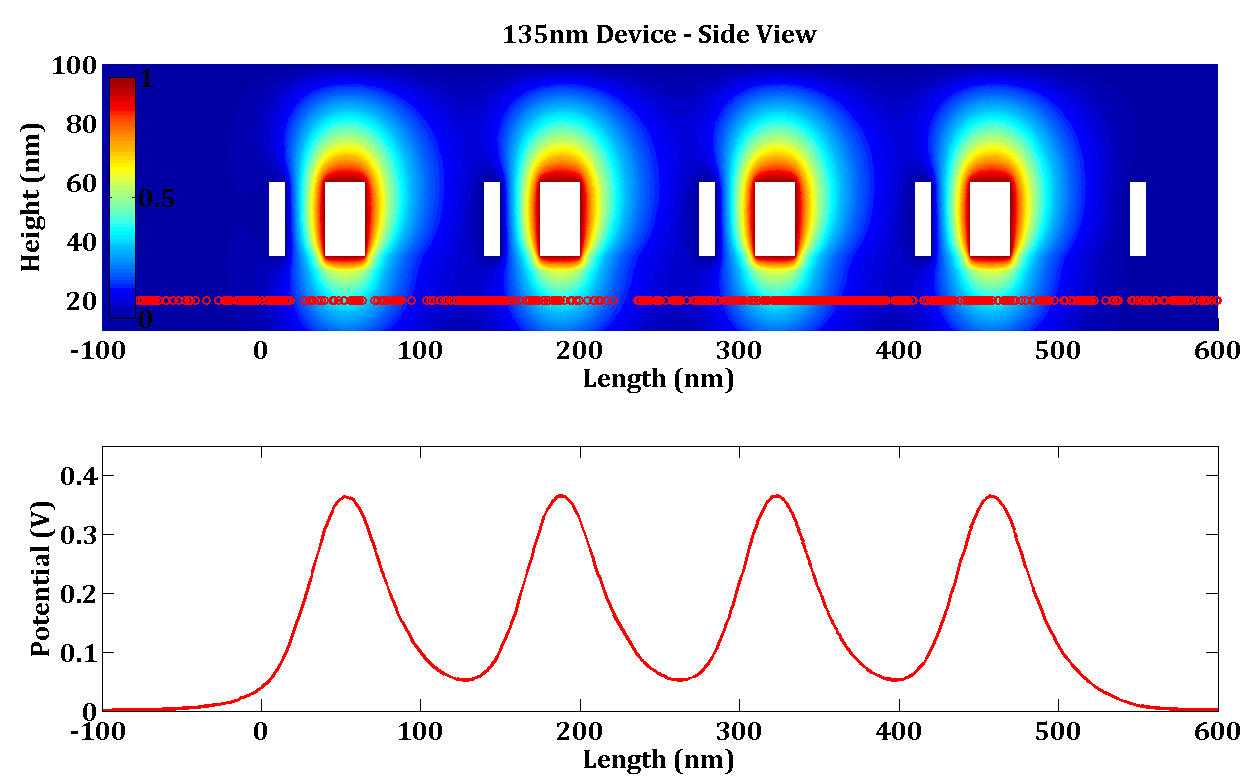}
\caption{Simulation of the potential profile for a ratchet with 135nm barrier length using the Laplace tool box in Matlab (PDEtool). The positive electrodes are 25nm in width whereas the grounded electrodes are 10nm in width to increase the asymmetry. The a/b separation factor between the electrodes is 3. The lower figure is showing the Laplace potential through the center of the Si channel.}
\label{fig:digitated}
\end{figure}

Figure \ref{fig:fig1} shows a possible implementation of a ratchet based switch. We have a dual gated device with a top and a back gate capacitor. The top gate, consisting of interdigitated metal plates, creates a spatially asymmetric potential - obtained by solving Laplace's equation (Fig. \ref{fig:digitated}). An AC signal applied to these plates creates a clock that raises and lowers the potential barriers periodically. The ratchet mechanism (described in the next section) creates a net non-zero DC current averaged over space and time, and this current progressively builds up a charge on the back gate capacitor. The back gate capacitor represents the input gate to the next ratchet in series (not shown), moving its Fermi energy and turning it on so that the second ratchet can now start shuttling charges. A suitable layout of a ratchet array can then realize a NAND, a NOR or other generic Boolean logic gates.

\section{Ratchet Physics: A Toy Model}
\begin{figure} [t]
\centering
\includegraphics[scale=0.45]{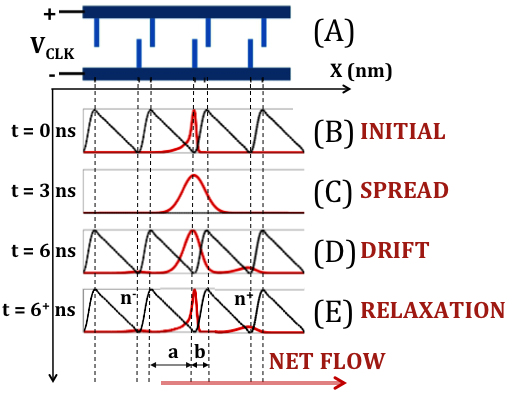}
\caption{Computing mechanism for a bistate quantum ratchet simulated with 135nm barrier length, $1.67 GHz$ AC clock, and $\sim20kT$ barrier height using {\bf quantum} flow and relaxation. {\bf (A)}: Interdigitated electrodes between two conductor planes creating the desired asymmetric local potential profile in the channel. {\bf (B)}: Initial carrier distribution (shaded area) when potential (dashed line) is ``on". {\bf (C)}: Distributed carriers when potential is ``off''.{\bf (D)}: electron distribution when potential is again turned back on at the heated state. The red line $n^-$ stands for the carriers overcoming the barrier peak on the left ({\bf back-flow}), while $n^+$ stands for those on the right ({\bf forward-flow}). Since $n^+$ is bigger than $n^-$, a net flow to the right results. {\bf (E)} Shows the relaxed stage at the end of each cycle, where we reset the electron distribution (n/n0) to the initial equilibrium solution.}
\label{fig:ratchet_basic}
\end{figure}

We consider only a specific subset of all possible ratchet types - `flashing ratchets'. Fig. \ref{fig:ratchet_basic} shows the basic operation of a flashing ratchet and how the particles will accumulate around a single potential minimum. However in a real device the particles will be distributed equally to every minima. A periodic potential with built-in local asymmetry is created by a sequence of interdigitated electrodes (Fig. \ref{fig:digitated}). When the barriers are fully raised, the carriers injected into the channel localize around the potential minima (Fig. \ref{fig:ratchet_basic}B). When the applied AC clock lowers the barriers (Fig. \ref{fig:ratchet_basic}C), the carriers spread out in both left and right directions, driven by thermal diffusion in the classical limit and by wave packet evolution in the quantum limit, driven by the difference in phase velocities of the individual Fourier components. When the potential is turned back on again, the carriers drift down to their nearest valleys (Fig. \ref{fig:ratchet_basic}D). Owing to the local asymmetry of the potential profile, there will be more carriers that have crossed the top of the barriers into each valley compared to the barrier sitting further, because the diffusion occurs when the barriers are down. In other words, there is a progressive space-time averaged movement of charges unidirectionally {\it{even in the absence of a source-drain bias}}, driven simply by the non-equilibrium signal supplied by the clock. 
 
The action of a ratchet can thus be described as a clock-driven current, that 
can be used to build an open-circuit voltage  $V_{OC}$  across a capacitor in a logic circuit. As the voltage builds on the capacitor, it creates a back-flow until the reverse bias current cancels the ratchet current upon complete charging of the capacitive load. A large open circuit voltage allows the ratchet to drive the next stage, but this efficiency needs to be counter weighed against the corresponding energy dissipated and the charging delay associated with building the voltage. 

\subsection{Approximate quasi-analytical result}
Using the diffusion equation \cite{pierret}, near the valley at $x_0$, the potential profile can be approximated as 
\begin{equation}
U(x) \approx U(x_0) + \frac{(x-x_0)^2}{2} U^{\prime \prime}(x_0)
\end{equation}
where dash represents a 1-D spatial derivative. The corresponding equilibrium initial 
carrier distribution becomes
\begin{eqnarray}
N(x) &\propto& \exp[-U(x)/kT] \nonumber \\
&=& n_0\exp\left[-\frac{(x-x_0)^2}{2 \sigma_0^2}\right]
\end{eqnarray}
where $\ds \sigma_0 = \sqrt{{kT}/{U^{\prime \prime}(x_0)}}$ is the standard deviation of Gaussian distribution and $U(x_0)$ is set to be zero for convenience. After one computing cycle (Fig. \ref{fig:ratchet_basic}), the net carrier density difference between two neighboring potential wells is 
\begin{eqnarray}
\Delta N &=& n_0 Sl_{barr} \Bigg[ erfc \left(\frac{b}{\sqrt{2 \sigma_0^2 +4Dt_{off}}}\right) \nonumber \\
&-&erfc \left(\frac{a}{\sqrt{2 \sigma_0^2 +4Dt_{off}}} \right) \Bigg]
\end{eqnarray}
where $erfc$ is the complementary error function, and $n_0 = N_c exp(-\frac{E_f - E_c}{kT})$ is the initial carrier density in the semiconducting channel, $N_c$ is the effective density of states, $E_f$ is the fermi-level, $E_c$ is the conduction band level, $D$ is the diffusion constant, $S$ is the channel cross-sectional area, $l_{barr}$ is the length of the potential barrier, $t_{on}$ is the duration of the ``on'' part of the clock when the barrier is raised, $t_{off}$ is the ``off'' time period when the barrier is lowered, $a$ and $b$ are the asymmetric physical lengths defined in Fig. \ref{fig:ratchet_basic}D. This expression also assumes that the drift time is smaller than $t_{on}$, i.e. the barrier is raised long enough for the charges to drift into the valley, whereupon phonons rapidly relax their energy. Finally, the net space- and time-averaged current under zero reverse bias becomes 
\begin{equation}
I_0 = \frac{q \Delta N}{t_{on} + t_{off}} 
\label{eqn:currenT}
\end{equation}
It is clear that the current arises because of the asymmetry ($a > b$), driven by the clock
frequency related to $t_{on,off}$. 

\begin{figure*} [t]
\centering
\includegraphics[scale=0.65]{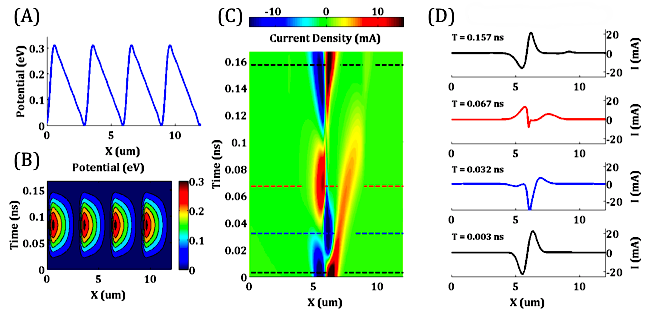}
\caption{{\bf (A)} shows a spatio-temporal ratchet potential, with a prominent spatial asymmetry with 4 barriers. {\bf (B)} shows the contour-plot of this ratchet potential progressively turned on between 0.02 ns and 0.08 ns and then turned off after 0.8 ns. {\bf (C)} shows a colormap of the current density over this time (blue is negative, red is positive, and green is zero). As seen from the figure, the drift segment of the current during the potential up-cycle is actually two-phased, showing a quick initial drift down the steeper slope (thin, localized dark blue to cyan), followed by a slower drift of a larger number of particles (hence a net positive current) down the smaller slope shown in spreading red to yellow. The current density plots shown in {\bf (D)} are the snapshots at time instances marked by the horizontal dashed lines in {\bf (C)}. During down cycle, the back and forward propagating currents (blue and red, near t = 0.157 ns) are comparable as the diffusion process is symmetric in space whereas when potential is turning up, near t = 0.032 ns back-flow dominates and at t = 0.067 ns the forward-flow is more prominent. Thus the net current flow proceeds with a two-phased drift, supplanted gradually by a symmetric diffusion.}
\label{fig:spread}
\vspace{-.1cm}
\end{figure*}

\subsection{Simulating a quantum ratchet} 
The quantum flow of electrons in the ratchet involves solving the time-dependent one-electron
Schr\"odinger equation 
\begin{equation}
i \hbar \frac{\partial \psi}{\partial t} - H \psi = 0
\label{eqn:schrodinger}
\end{equation}
where $H$ is the time-dependent Hamiltonian matrix describing the channel. The Hamiltonian is described using the 1-D finite difference tight-binding formula 
\begin{equation}
H_{n,m} = [U_n + 2 t_0] \delta_{n,m} - t_0 \delta_{n,m+1} - t_0 \delta_{n,m-1}
\label{eqn:hamiltonian}
\end{equation}
where $t_0 \equiv \hbar^2 / 2 m^{\ast} a^2$ depends on the grid size $a$ and the effective mass $m^{\ast}$. To get ballistic non-equilibrium flow in a 
drain-driven device, we normally add in self-energy matrices for injection and removal, coupled with bias-separated contact Fermi-Dirac distributions \cite{dattabook}. For the ratchet however, flow is generated {\it{simply by the time-dependence in $H$}}. To visualize this short-circuit current, we will simply incorporate periodic boundary conditions for the shuttling of charges. The initial state of the particles is obtained from the eigenvectors $\{\alpha\}$ and eigenvalues $E_{\alpha}$ of the matrix $[H]$ at the time instant $t=0$, while the subsequent evolution of the wavefunctions is obtained using the Crank-Nicholson approximation for its efficiency and simplicity in computing :
\begin{equation}
\psi_{t + \Delta t} = [1 - \frac{i H \Delta t} {2 \hbar}]^{-1} [1 + \frac{i H \Delta t} {2 \hbar}] \psi_t
\label{eqn:crank}
\end{equation}

In addition to a local asymmetry and an energy source, the ratchet needs dissipation to generate a `reset' at the end of each cycle. A purely ballistic quantum evolution, described above, does not have an inbuilt mechanism for relaxing the charges. As a result, our simulations show that the process of continuously pumping energy into the ratchet from an AC field causes the carriers to heat up and eventually fly off the barriers. We thus need a way to remove the excess energy at the end of each AC cycle. Coupling the electrons with substrate phonons would remove that energy. Rather than modeling this complex behavior, we capture this relaxation using a `one-shot' procedure, where at the end of each clock cycle, we reset the electron distribution to the initial equilibrium solution. The energy difference between the equilibrium eigenvalue and that reached at the end of the AC cycle is the dissipation that we can then keep track of. 

The time-dependent potential $U_n \equiv U(x_n)$ samples the asymmetric barrier shape, which we choose as $U(x) = x^2 (1-x^7)$. Periodic boundary conditions are invoked by setting $H_{1,N} = H_{N,1} = -t_0$, while open boundary conditions require self-energy matrices with $\Sigma_1(1,1) = \Sigma_2(N,N) = -t_0e^{ika}$. While periodic boundary conditions suffice to see the shuttling of charges, and in addition provide analytical simplicity for a closed circuit, there is a danger of tails of the wavefunction escaping a drain and re-injected from the source towards a valley and self-interfering, an issue that open boundary conditions do not run into. This becomes a bigger issue if we let the charges build up at the end for an open circuit, whereupon the periodic boundary conditions cannot even be justified. Consequently, we will use open boundary conditions for the latter half of the paper (see section III-E).

\subsection{Simulating a classical ratchet}
\begin{figure} [t]
\centering
\includegraphics[scale=0.38]{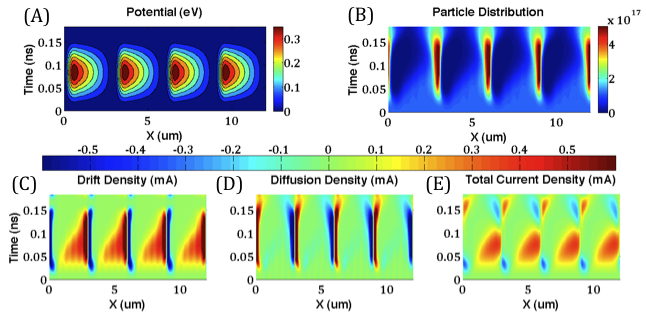}
\caption{{\bf (A)} above shows how the asymmetric ratchet potential changes with time and space. {\bf (B)} shows how the particles are evenly distributed (localized) in bins when the potential is raised and how they move to the right with the oscillating top gate (the blue area in between localized bins shows the pathway of particles moving right). {\bf (C-E)} show the total, drift, and diffusion current densities. One can see that drift current opposes the diffusion current at all the times, but their magnitudes are different thus resulting in a net flow to the right. As discussed in the previous sections, the second phase of the two-part drift current shown in yellow moves the particles to right, then the diffusion current spreads the particles evenly right and left.}
\label{fig:JXT}
\end{figure}

Classically we keep track of the charges rather than wavefunctions, and evolve them by the Newtonian drift-diffusion equation
\begin{equation}
\frac{\partial N}{\partial t} = -\frac{\partial (\mu \xi N)}{\partial x} + D \frac{\partial^2 N}{\partial x^2}
\label{eqn:carrier}
\end{equation}
where $N=N(x,t)$ is the carrier distribution, $\xi=\xi(x,t)$ is the electric field, $\mu$ is the carrier mobility, and D is the diffusion constant. The triangular potential affects the carriers through the electric field given by the equation:
\begin{equation}
\xi = -\frac{dU}{dx}
\end{equation}
where $U$ is the spatially-varying potential profile across the ratchet. The current can be calculated from N as
\begin{equation}
J = qvN + qD\frac{\partial N}{\partial x}
\label{eqn:current}
\end{equation}
where $v=\mu \xi$ and $q$ is the electric charge. Fig. \ref{fig:spread}A and B shows how the potential is varied with space and time. Part C demonstrates how the total current density is generated by the movement of electrons and has a net positive current flow to the right. The Gaussian distributed particles initially localize in the second well (L $\approx 6\mu m$) and start to diffuse both sides at t = 0.003 ns equally. As the potential starts to turn on between 0.02 ns and 0.08 ns, a small amount of particles (due to asymmetry of potential) quickly drift down the steeper slope (thin dark blue to cyan) followed by a slower, positive drift movement of a larger group of particles (red/orange to yellow) resulting in a net positive forward-flow. At t = 0.157 ns, the potential is lowered down and the diffusion starts, helping the two-phase drift current. As shown in the current figures of Fig. \ref{fig:JXT}E, the net flow of particles is caused mostly by the drift current (Fig. \ref{fig:JXT}C), not by the diffusive current which is almost symmetrical in both directions (Fig. \ref{fig:JXT}D). 

The boundary condition at the contacts is set to describe carrier recombination in the relaxation time approximation
\begin{equation}
\frac{dN}{dt} \bigg|_{end} = -\frac{N_{end}}{\tau}
\end{equation}
where $\tau$ is the time it takes for the electrons to transfer from the semiconductor to the capacitor. In these simulations, we use copper as the capacitive material for the back gate of the next ratchet (See Table \ref{table1}) \cite{knorren}. This boundary condition occurs because of the interface between the ratchet channel and contact, and it usually limits the amount of charge which can flow in and out of the capacitor. However, in our simulations, the charge extraction rate for copper was fast enough to allow all charge carriers to (dis)charge the capacitor. Therefore, the contact interface did not limit the current flow between the capacitor and the ratchet. 

In order to capture the charging process of the load capacitor, the number of charges accumulating on the capacitor was calculated as $Q = J_{end} S \Delta t$, where $J_{end}$ represents the current density crossing the boundary at the end of each time step, $S$ is the channel cross sectional area, and $\Delta t$ is the simulation time step. Using this we can calculate the voltage increase as $V = {Q}/{C}$, where $C$ is the load capacitance. We then superimpose a backward voltage varying linearly from 0 to $V$ across the ratchet (ideally, by solving Poisson's equation, but simplified in our treatment here), and add it to the sawtooth potential to give a tilted asymmetric potential. We then recalculate the electric field and reiterate eqn. \ref{eqn:carrier} until the capacitor is charged under open bias and the net {\it{space-time-averaged current}} vanishes (shown in Fig. \ref{fig:capcharge}). Note that there is still dynamic current flow and associated dissipation. 

\subsection{Calculating the Open Circuit Voltage}
Charging of the load capacitor increases the voltage on one side of the ratchet, resulting in a tilted potential growing with time across the device. The resulting reverse current  partially cancels out the directional current from the ratchet. Eventually, the capacitor reaches a critical voltage, called the open circuit voltage ($V_{OC}$) when the reverse current equals the current generated by the ratchet, averaged over space and time. It is therefore important to extract the open circuit voltage to understand the drivability of the ratchet device for logic. 

\begin{table}[t]
\renewcommand{\arraystretch}{1.3}
\caption{Simulation Parameters for Drift-Diffusion \newline Refer to Table II for Circuit Parameters}
\label{table1}
\centering
\begin{tabular}{| c | c |}
\hline
Barrier Length ($L$) & $3~\mu m$ \\
Asymmetry ($a$ - $b$) & $2.25~\mu - 0.75~\mu m$ \\
Channel Thickness ($t$) & $20~nm$ \\
Channel Width ($W$) & $6~\mu m$ \\
Oxide Thickness ($t_{ox}$) & $5~nm$ \\
Top Gate AC Clock & 0.39V @ 6 GHz \\
Diffusion Constant ($D$) & $3.62\times10^{-3}~m^2 s^{-1}$ \\
Si Eff. Den. of States ($N_c$) & $3.23\times10^{25}~m^{-3}$ \\ 
$\eta = (\frac{E_c - E_f}{kT})$ & 3 \\ 
Max. Ratchet Barrier Height (in $kT/q$) & 15 \\
Capacitor Escape Rate (for Cu) ($\tau$) & $1.786\times10^{-14}~s$ \\
\hline
\end{tabular}
\end{table}

\begin{figure*} [t]
\centering
\includegraphics[scale=0.80]{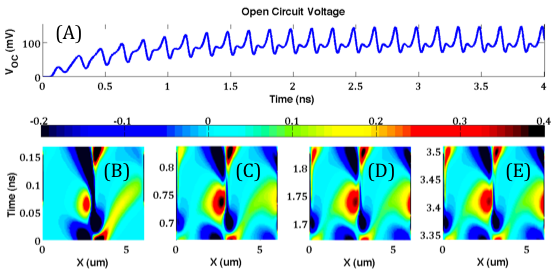}
\caption{{\bf (A)}Drift-diffusion simulation of the electronic ratchet showing the charging of a load capacitor. After 11 clock cycles the particles are distributed equally among all the barriers and there is a net forward-flow of current to the right. The two-phase drift current is less prominent as the $V_{OC}$ is built-up, but still exists as seen in the {\bf (B-E)} (Darker red colors start changing to orange/yellow colors as the $V_{OC}$ increases). As the particles reach the right side (L = 6$\mu$m), they get ejected by the system and charge a localized back gate capacitor to a $V_{OC}$ of $\approx145mV$.}
\label{fig:capcharge}
\end{figure*}

Let us assume that the time dependent current through the ratchet in one operation cycle can be given by $I(t)$. Therefore, the current generated by the ratchet after one clock cycle is 
\begin{align}
I_{0} &= \frac{1}{t_{on}} \int_{t_{off}} ^{t_{on}+t_{off}} {I(t)}dt \notag\\
&= \frac{Q}{t_{on}}
\end{align}
where $Q$ is the total charge placed on the load capacitor. The voltage generated on the load capacitor after one cycle is $V_{0} = {Q}/{C}$. This, in turn, creates a reverse current of $I_{rev,0}={V_{0}}/{R}$, where $R$ is the resistance of the ratchet channel (in general, a nonlinear resistance which must be calculated self-consistently). If we assume that the space-time averaged current pumped by the ratchet remains constant, the total current generated by the ratchet during the next cycle given by
\begin{align}
I_{1} = \frac{Q}{t_{on}} - I_{rev,0}
\end{align}
This way, the voltage on the capacitor after $N$ cycles is
\begin{align}
V_{N} &= \frac{1}{C} I_{N} t_{on} + V_{N-1} \notag \\
&= \frac{1}{C}\left [\frac{Q}{t_{on}} - \frac{V_{N-1}}{R}\right] t_{on} + V_{N-1} \notag \\
&=\frac{Q}{C} + (1-\frac{t_{on}}{RC}) V_{N-1} \notag \\
&=\frac{Q}{C}\left[\frac{1-(1-\displaystyle\frac{t_{on}}{RC})^N}{\displaystyle\frac{t_{on}}{RC}}\right]
\end{align}
Therefore, the open circuit voltage can be defined as
\begin{align}
V_{OC} &\equiv \lim_{N \to \infty} V_{N}  = \frac{Q}{t_{on}} R \notag \\
&= \frac{l_{barr}}{q n_0 \mu S} \cdot \frac{1}{t_{on}}\int_{t_{off}} ^{t_{on}+t_{off}} {I(t)}dt
\label{eqn:voltage}
\end{align}
It is important to note that the output voltage is only \emph{indirectly} dependent on the input voltage. The input voltage from the back gate increases the number of carriers in the channel, and thus the magnitude of $V_{OC}$. However, since the top gate is always oscillating, it does not influence the input that tries to determine the output. \emph{The input is decoupled from the charging process of the output which means that the ratchet back gate is adiabatically charged without placing any timing constraints on the input}. 

If we now apply Eqn. (\ref{eqn:voltage}) to (\ref{eqn:currenT}), we get the following $V_{OC}$ for the Gaussian distribution
\begin{align}
V_{OC} &= \frac{l_{barr}^2}{2 \mu (t_{on} + t_{off})} \Bigg[ erfc \left(\frac{b}{\sqrt{2 \sigma_0 + 4 D t_{off}}} \right) \notag \\
&- erfc \left(\frac{a}{\sqrt{2 \sigma_0 + 4 D t_{off}}} \right) \Bigg]
\label{eqn:voltage2}
\end{align}

For the simulations, we use the parameters listed in Table \ref{table1}. The parameters are conventional material properties of silicon for the channel and silicon dioxide for the oxide \cite{pierret}. Furthermore, we derive device parameters such as back gate capacitance from the dimensions and structures shown in Fig. \ref{fig:fig1}.

In figure \ref{fig:capcharge}, we show the charging of the load capacitor. According to the drift-diffusion simulations, the open circuit voltage saturates to $\approx 145 mV$, which is in a relatively good agreement with the analytic value of $\approx 125 mV$ derived by using the parameters from Table \ref{table1} in eqn. (\ref{eqn:voltage2}). From 3D figure showing different time cycles in Fig. \ref{fig:capcharge}A, we can see that as the forward-flow of electrons decreases with time and $V_{OC}$ builds up. In the Fig. \ref{fig:capcharge}B at t $\approx 0.05 ns$ and x $\approx 5 \mu$m, there is net positive current shown in yellow, but as the $V_{OC}$ built, this positive current gets lets prominent and disappears as shown in Fig. \ref{fig:capcharge}E. The space-time averaged net-current becomes zero. 

\section{Logic with Ratchets}
\subsection{Static power in CMOS vs. Ratchet}
\begin{figure} [t]
\centering
\includegraphics[scale=0.20]{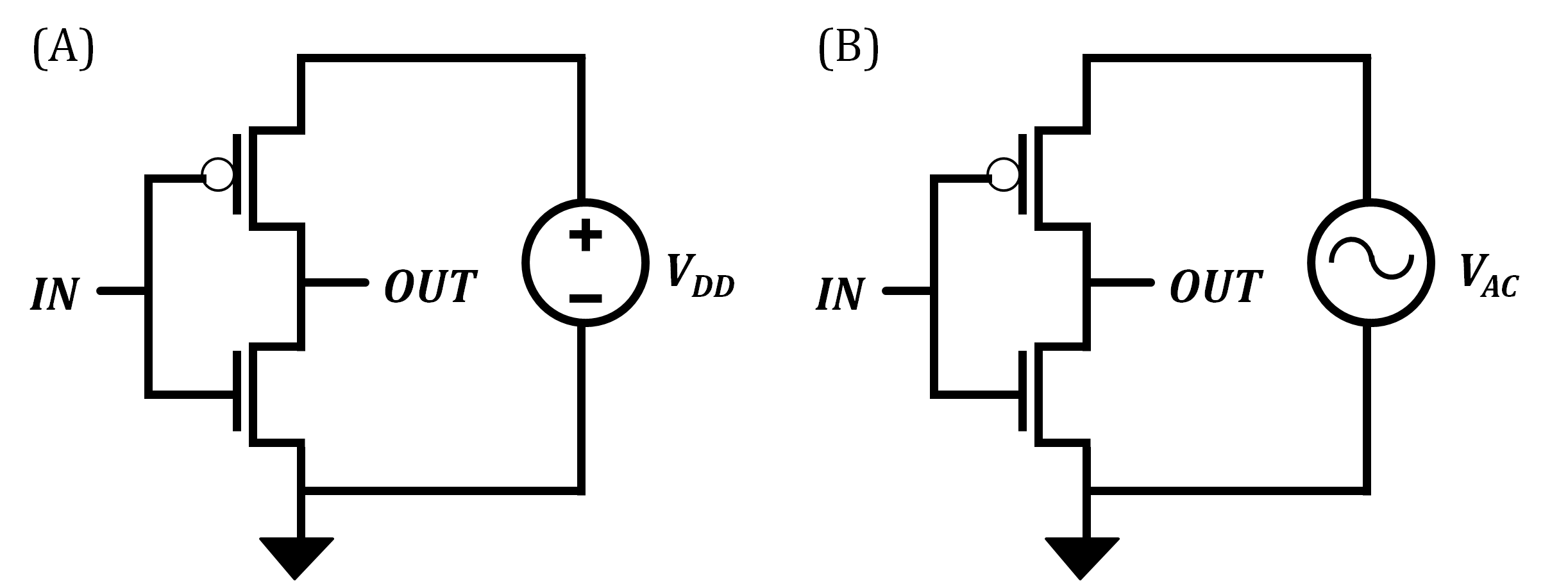}
\caption{Conventional vs. Adiabatic Charging \textbf{(A)} CMOS circuits are conventionally charged using a DC voltage source. \textbf{(B)} By using a low frequency AC voltage source, the capacitor can be charged while dissipating less energy \cite{tnano11}. }
\label{fig:fig2}
\end{figure}

Static power dissipation in CMOS circuits is caused by a direct current from the power source ($V_{dd}$) to ground (See Fig. \ref{fig:fig2}A). In an ideal CMOS circuit, there is no static dissipation because its complementary nature ensures that for each turned on PMOS, there exists a turned off NMOS blocking a direct path to ground, or vice versa. However, in real circuits, MOSFETs are not completely turned off and there exist leakage currents which traverse from $V_{dd}$ to ground--creating static power dissipation. In modern CMOS circuits, leakage power is a significant problem which usually accounts for a large portion of all power dissipation \cite{yeo}.

The electronic ratchet, on the other hand, can avoid this issue altogether. Rather than using a DC power source to generate a current, the ratchet uses a time varying asymmetric potential to create a net current. Therefore, \emph{the ratchet can drive current without a source to drain bias}. Since there are no DC power sources, the ratchet can, in principle, be used to design circuits with no static dissipation. 

\begin{table}[b]
\renewcommand{\arraystretch}{1.3}
\caption{Parameters for Circuit Simulations \newline Refer to Table I for device dimensions}
\label{table2}
\centering
\begin{tabular}{| c | c |}
\hline
Freq. of Top Gate ($\omega$) & $6\times10^9~s^{-1}$ \\
Top Gate Capacitance ($C_1$) & $9.07\times10^{-13}~F$ \\
Back Gate Capacitance ($C_2$) & $4.24\times10^{-13}~F$ \\
Inductance ($L$) & $7.76\times10^{-10}~H$ \\ 
Channel Resistance ($R$) & $1.39\times10^{3}~\Omega$ \\
\hline
\end{tabular}
\end{table}

\subsection{Dynamic power in CMOS vs. Ratchet}
In CMOS logic circuits, dynamic power dissipation results from the charging and discharging of load capacitors. Take the inverter in Fig. \ref{fig:fig2}A for example. When the inverter has an input of '0', the load capacitor is charged to $Q = CV_{dd}$. Therefore, the energy dissipated is given by:
\begin{eqnarray}
E_{diss} &=& E_{source} - E_{cap} \nonumber \\
&=& \int_0^{Q}V_{dd}~dQ - \int_0^{Q}V~dQ \nonumber \\
&=& CV_{dd}^2 - C\int_0^{V_{dd}}V~dV \nonumber \\
&=& \frac{1}{2} CV_{dd}^2
\end{eqnarray}
This energy is dissipated in the form of heat due to the effective resistance of system. The dissipated power is given by the equation $P = \frac{1}{2} CV_{dd}^2 f$, where $f$ is the operating frequency of the CMOS circuit. An alternative method of charging which reduces the dynamic dissipation is using a \emph{voltage-controlled current source} like the electronic ratchet. It will be shown in Section III-C that a load capacitor charged through a voltage-controlled current source is:
\begin{equation}
E_{ratchet} = \frac{1}{2} C V_{OC}^2 \left(\frac{R}{R+{V_{OC}/{I_m}}}\right) 
\end{equation}
where $I_m$ is the maximum current through the ratchet when the load capacitor is uncharged (i.e., the short-circuit current), $V_{OC}$ is the final open circuit voltage across the ratchet, and R is the external interconnect resistance. While $V_{OC}$ effectively plays the role of a drain bias that develops across the ratchet, with a magnitude that is set by the desired ON-OFF ratio, the added term in the denominator helps shave off some of the dynamic power dissipation.

\begin{figure} [t]
\centering
\includegraphics[scale=0.4]{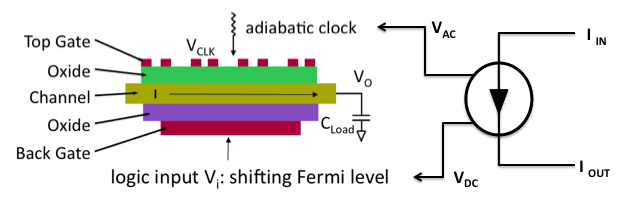}
\caption{Structure of a single ratchet and its circuit representation, showing the input and output ports of the ratchet modeled a voltage-controlled current source.}
\label{fig:symbol}
\end{figure}

Another way to further reduce this dissipated energy is a method known as adiabatic charging \cite{denker}. Fig. \ref{fig:fig2}B shows an example of an adiabatic CMOS circuit, where the DC power source has been replaced by an oscillating signal, known as a clock. Now let's assume that the input is '0' as the clock begins to ramp up. If it takes $t=\Delta T$ for the load capacitor to charge to $V_{dd}$, then the energy dissipated during the charging process can be given as:
\begin{eqnarray}
E_{diss} &=& \eta P \Delta T \nonumber \\
&=& \eta I^2 R \Delta T \nonumber \\
&=& \eta \left(\frac{CV_{dd}}{\Delta T}\right)^2 R \Delta T, 
\end{eqnarray}
where $\eta$ is a factor which is dependent on the shape of the waveform of the clock \cite{takashi}. Notice that by increasing $\Delta T$, we can reduce the dissipated energy indefinitely. However, one of the drawbacks of adiabatic CMOS is that it requires timing information for the inputs in order to maintain adiabatic operation. For example, it is important that we turn on the transistor when the clock is ramping up. If we had instead turned on the transistor when the clock was at its maximum voltage, then there would have been a large voltage drop across the effective resistance of the circuit, leading to non-adiabatic dissipation.  

In the electronic ratchet, the time-varying asymmetric potential acts as the clock in an analogous fashion as the adiabatic CMOS. However, as we showed in Section II, \emph{the input of the ratchet is decoupled from the charging process of the output}. This allows us to take advantage of the adiabatic charging without imposing timing constraints on the inputs. In Fig. \ref{fig:symbol}, a circuit symbol of electronic ratchet is presented showing the input and output ports.

\subsection{Capacitive Charging using a Current Source}
\begin{figure} [t]
\centering
\includegraphics[scale=0.4]{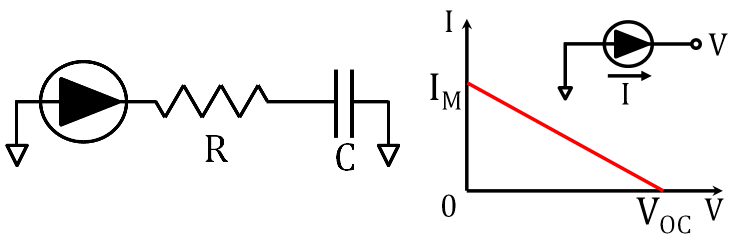}
\caption{An equivalent circuit of the charging behavior of the ratchet. The ratchet acts as a voltage-controlled current source, where current is a function of the voltage on the capacitor. The current is at a maximum, $I_m$, initially when the capacitor is uncharged. As the voltage builds up, the reverse current decreases the net current, until the capacitor reaches the open circuit voltage, $V_{OC}$ when there is a zero net current.}
\label{fig:ratchetIV}
\end{figure}

Consider a simple circuit model in which an electronic ratchet charges a capacitor, as shown in Fig. \ref{fig:ratchetIV}. The I-V characteristic of the electronic ratchet is also shown.
We can write the Kirchoff's equations as follows to describe this circuit
\begin{eqnarray}
&I& = \frac{V-V_c}{R} \\
\label{eqn:resistor}
&I& = -\frac{I_m}{V_{OC}}V + I_m \\
\label{eqn:currentsource}
&I& = C \frac{dV_c}{dt} 
\label{eqn:capacitor}
\end{eqnarray}
where $I$ is the current, $V$ is the voltage of the node between the electronic ratchet and the resistor and $V_c$ is the voltage on the capacitor.

From the equations (\ref{eqn:resistor}-\ref{eqn:capacitor}), the voltage on the capacitor, $V_c$, and current, $I$ can be
solved as
\begin{eqnarray}
V_c &=& V_{OC} (1-e^{-{t}/{R^{\prime}C}}) \nonumber \\
\label{eqn:Vc}
I &=& \frac{V_{OC}}{R^{\prime}} e^{-{t}/{R^{\prime}C}}
\label{eqn:I}
\end{eqnarray}
in which $\ds R^{\prime} = R + {V_{OC}}/{I_m}$

Therefore, the energy charged into the capacitor is
\begin{equation}
E_{cap} = \int_0^{V_{OC}} C V_c dV_c = \frac{1}{2} CV_{OC}^2
\label{eqn:Ecap}
\end{equation}
and the dissipated energy on the resistor is 
\begin{eqnarray}
E_R &=& \int_0^{\infty} I^2 R dt \nonumber \\
&=& \int_0^{\infty} (\frac{V_{OC}}{R^{\prime}})^2 e^{-{2t}/{R^{\prime}C}} R dt \nonumber \\
&=& \frac{CV_{OC}^2}{2} \frac{R}{R^{\prime}} \nonumber \\
&=& \frac{CV_{OC}^2}{2} \frac{R}{R+\frac{V_{OC}}{I_m}}
\label{eqn:ER}
\end{eqnarray}
From this equation, we make the following observations. In a ratchet, the load capacitor is charged to $V_{OC}$ rather than $V_{dd}$. Therefore, the ratchet energetics are comparable to CMOS circuits with extremely low drain-source voltages. Furthermore, the ratchet introduces an extra resistance term, ${V_{OC}}/{{I_m}}$ which decreases the energy dissipation further. For our simulations (see Section 3), the channel resistance was 1.39 k$\Omega$ compared to the extra resistance term of 2.70 k$\Omega$. Therefore, the unique charging method of the ratchet makes it intrinsically less dissipative than conventional CMOS logic.

\subsection{Adiabatic Charging}
\begin{figure} [t]
\centering
\includegraphics[scale=0.4]{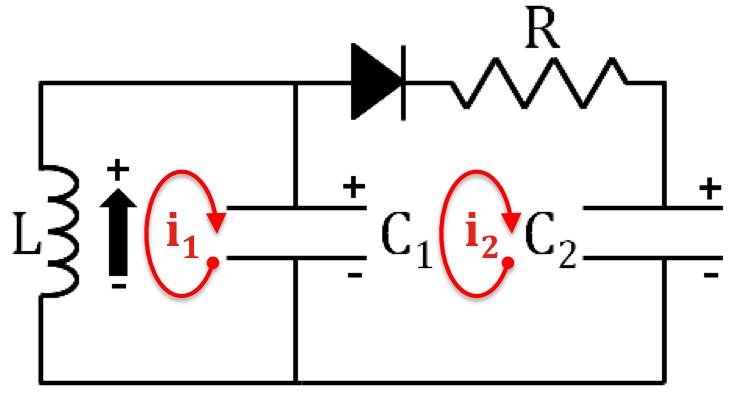}
\caption{Circuit model for the ratchet. An adiabatic clock is modeled using an inductor \textbf{L} and the top gate capacitor \textbf{C1}. A (virtual) diode is used to capture the unidirectional current (ratchet) which drives charging of the load capacitor \textbf{C2}.}
\label{fig:circuit}
\end{figure}

We will now develop a more accurate circuit model for the ratchet (Fig. \ref{fig:circuit}) to describe its energetics during adiabatic charging.
An LC circuit is used to represent the oscillating clock for the top gate. This clock charges up the top gate capacitor, which represents the charging or discharging of the asymmetric ratchet potential. Because the circuit is operating at the natural frequency, we are able to adiabatically transport electrons in the channel, thus reducing the energy dissipation of the circuit. 

The asymmetry of the ratchet potential provides directionality to the electrons and gives a net positive current. This asymmetry is captured by a (virtual) Schottky diode, which allows current to flow easily in one direction while acting as a rectifier in the opposite direction. The diode parameters are selected by calibrating with the drift-diffusion simulation. The barrier height is selected as $15 kT/q$ as in the simulations and the $I_0$ is selected as 1x$10^{-15}A$ to match the results. The diode also acts as the voltage-controlled current source described in Fig. \ref{fig:ratchetIV}. This current is used to charge the load capacitor, which acts as the back gate capacitor of the next ratchet in a cascaded logic architecture (Fig.~\ref{fig:cascade_ratchet}). 

The charging process can be seen in the following analysis. We treat the diode as a short-circuited element when there is a positive current running through it, and as an open-circuit when the current is negative . During the time when the diode is short circuited, we have the following Kirchoff's equations
\begin{eqnarray}
\label{eqn:k1} 
L \frac{di_1}{dt} - \frac{Q_1}{C_1} = 0 \\ 
\frac{Q_1}{C_1} - Ri_2 - \frac{Q_2}{C_2} = 0
\label{eqn:k2}
\end{eqnarray}
where $i_1$ and $i_2$ are the currents in the two separate loop circuits. For eqn. (\ref{eqn:k1}), we assume that the charge on the top gate capacitor, $C_1$, is exclusively determined by the current generated by the inductor. This is a valid assumption if the top gate capacitance is much larger than the back gate capacitance (see Table \ref{table2}), thus acting as an `energy tank' \cite{blanchard1941history}. 

\begin{figure} [t]
\centering
\includegraphics[scale=0.55]{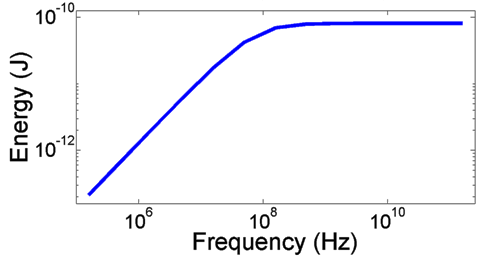}
\caption{Energy dissipated as a function of $\omega_L$ for a fixed input energy on the inductor. The energy dissipated is calculated by taking the integral of the dissipated power $I^2 R$ at $t=\infty$. The energy goes to zero in the adiabatic limit.}
\label{fig:energy}
\end{figure}

We first solve eqn. (\ref{eqn:k1}), and apply the boundary condition ${dQ_1}/{dt}|_{t=0}=-I_0$ 
\begin{eqnarray}
\frac{d^2 Q_1}{dt^2} - \frac{Q_1}{LC_1} = 0 \nonumber \\ 
Q_1(t) = -\frac{I_0}{\omega_L} \sin{\omega_L t}
\label{eqn:inter}
\end{eqnarray}
where $\omega_L = {1}/{\sqrt{LC_1}}$. Next, we substitute (\ref{eqn:inter}) into (\ref{eqn:k2}) to get:
\begin{equation}
\frac{dQ_2}{dt} + \frac{Q_2}{RC_2}  = -\frac{I_0}{\omega_L RC_1} \sin{\omega_L t}
\label{eqn:inter2}
\end{equation}
Let $\omega_1 = {1}/{RC_1}$ and $\omega_2 = {1}/{RC_2}$. Therefore, equation (\ref{eqn:inter2}) has the solution 
(with boundary condition $Q_2(0)=0$) :
\begin{eqnarray}
Q_2(t)  &=&-\frac{I_0 \omega_1}{\omega_L^2 + \omega_2^2}\ exp({-\omega_2 t})  \nonumber \\
&-& \frac{I_0 \omega_1}{\omega_L (\omega_L^2 + \omega_2^2)} (\omega_2 \sin{\omega_L t} - \omega_L \cos{\omega_L t})
\end{eqnarray}

\begin{figure} [b]
\centering
\includegraphics[scale=0.35]{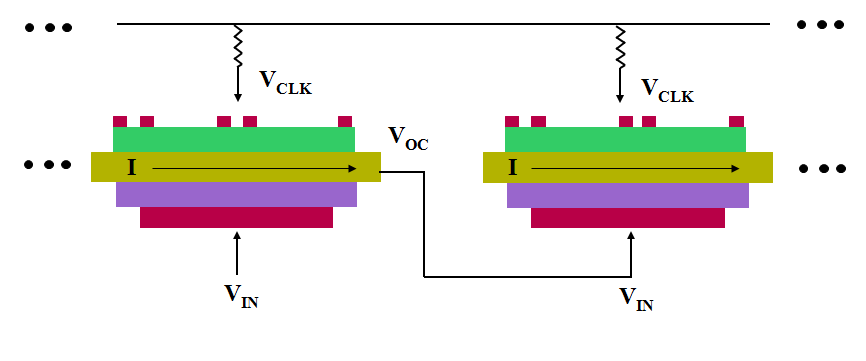}
\caption{Logic circuits are built by cascading ratchet gates together. The first ratchet builds up an open circuit voltage which is then used to charge or discharge the back gate of the next ratchet.}
\label{fig:cascade_ratchet}
\end{figure}

Therefore, the back gate charging current is 
\begin{eqnarray}
I(t) &=& \frac{I_0 \omega_1 \omega_2}{\omega_L^2 + \omega_2^2}\ exp({-\omega_2 t}) \nonumber \\
&-& \frac{I_0 \omega_1}{(\omega_L^2 + \omega_2^2)} (\omega_L \sin{\omega_L t} + \omega_2 \cos{\omega_L t})
\label{eqn:inter3}
\end{eqnarray}
Equation (\ref{eqn:inter3}) is only valid during times when $I(t) \geq 0$. This condition requires that:
\begin{equation}
\cos{\omega_L t}  \left( \frac{\omega_L}{\omega_2} \right) \sin{\omega_L t} \leq \exp{(-\omega_2 t)}
\label{eqn:inequality}
\end{equation}
Equation \ref{eqn:inequality} suggests that there are periods when the back gate capacitor stops charging. During these periods, the charge is maintained on the back gate capacitor, as we will see later in our numerical simulations. 

Let us now look at the adiabatic limit for the energy dissipation, by taking $\omega_L \rightarrow 0$. However, {\it{we must make sure that the initial energy on the inductor is unaffected by this limit}}. Since $\omega_L = 1/\sqrt{{L C_1}}$, L increases quadratically as we decrease $\omega_L$. Now, the initial energy on the inductor is given by $E_{ind} = \frac{1}{2} L I_{0}^2 \propto \left({I_0}/{\omega_L}\right)^2$. Therefore, we must decrease $\omega_L$ in such a way that {\it{${I_0}/{\omega_L}$ remains constant}}. It is easy to see that the second and third terms in equation (\ref{eqn:inter3}) go to zero as $\omega_L \to 0$. The first term also goes to zero if we re-write the equation as:
\begin{equation}
\lim_{\omega_L \to 0} I(t) = \frac{I_0 \omega_1 \omega_2}{\omega_L (\omega_L^2 + \omega_2^2)} \omega_L \exp(-\omega_2 t) = 0
\end{equation} 
With $I \to 0$, the dissipated energy through the resistor, $E_{diss} = I^2 R$ also goes to zero. This implies that the total energy dissipated can be reduced by arbitrarily lowering the clock frequency, as predicted by the adiabatic limit (Fig. \ref{fig:energy}). 

\begin{figure} [t]
\centering
\includegraphics[width=80mm]{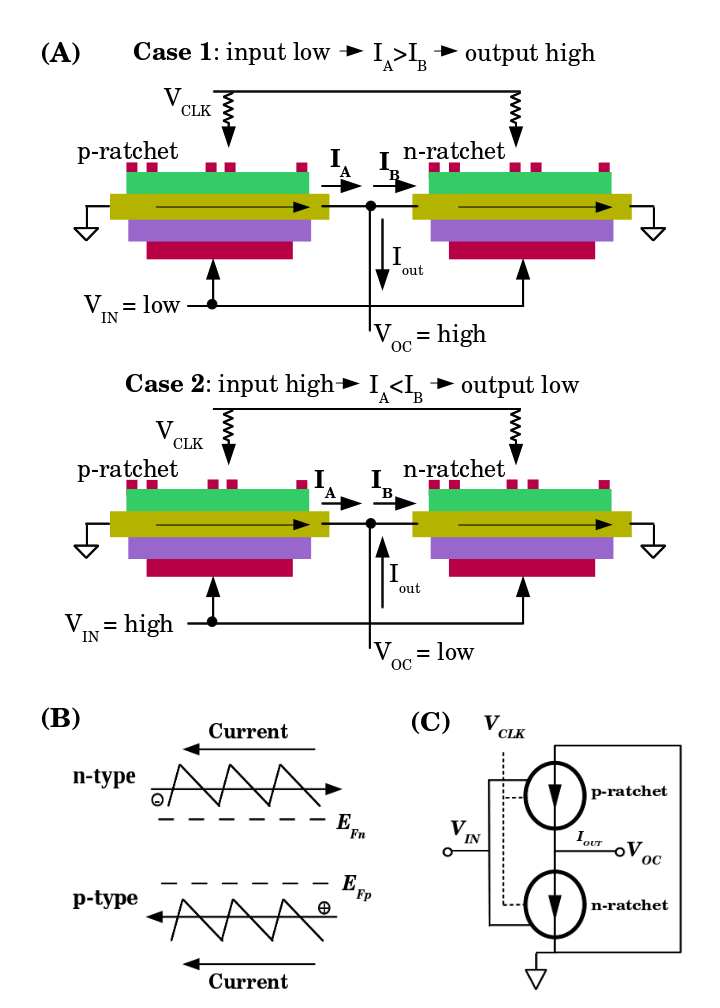}
\caption{Ratchet inverter computing mechanism. {\bf (A - Case 1)} $V_{in}$ is low and moves Fermi level down shown in {\bf (B)}, p-ratchet will have a higher current than n-ratchet so the output will charge the next capacitor $C_{bg}$ until the capacitor voltage reaches $-V_{OC}$ (low). $V_{in}$ is high the output will discharge next capacitor $C_{bg}$ from $V_{OC}$(high) to zero {\bf (A - Case 2)}.}
\label{fig:ratchet_inverter}
\end{figure}

\subsection{Ratchet-Based Gates}
The open circuit voltage generated across a ratchet can be used to move the Fermi energy across another ratchet, i.e., to electrostatically `dope' it (Fig.~\ref{fig:cascade_ratchet}). Electronic ratchets can be doped to either n-type or p-type, as shown in the Fig. \ref{fig:ratchet_inverter}B, by charging the back gate capacitor. CMOS-like combinations of an n-ratchet and a p-ratchet can further create universal ratchet based logic gates like INV, NAND, and NOR.  In theory, electronic ratchets are capable of realizing any Boolean logic operation based on these three gates. For example, a ratchet inverter can be realized with an n-type and p-type ratchet, as shown in Fig. \ref{fig:ratchet_inverter}A. As seen in the figure, the magnitudes of $I_A$ and $I_B$ (currents from the p-type and n-type ratchets) can be manipulated through the back gate, which adjusts the Fermi level in the channel (Fig. \ref{fig:ratchet_inverter}B). Furthermore, the output $V_{OC}$ is used to drive the back gate of the next ratchet in the circuit (shown as $V_{OC}$ in Fig. \ref{fig:ratchet_inverter}C). The sign of the current difference between $I_A$ and $I_B$ determines the charging or discharging of the next ratchet's back gate capacitor. When the input is high, the Fermi-level of the n-type devices gets shifted into the conduction band, allowing a large number of electrons to conduct and generate a high magnitude current ($I_B$). In the p-type ratchet, high input bias shifts the $E_F$ away from the conduction band, thus lowering the current ($I_A$). When $I_B$ is larger than $I_A$, the output of the inverter will start to discharge the next logic element's back gate capacitor.

\begin{figure} [t]
\centering
\includegraphics[scale=0.23]{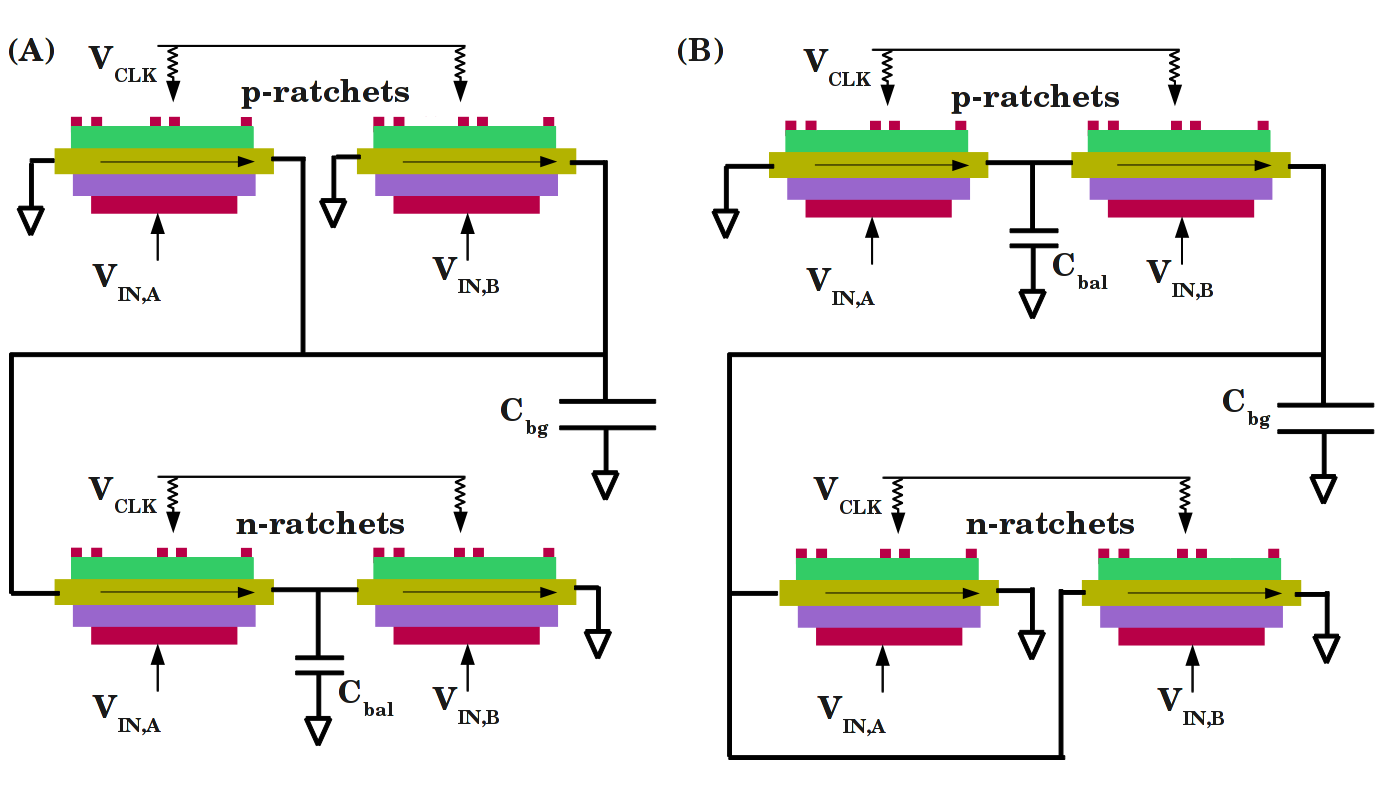}
\caption{{\bf (A)} Ratchet NAND gate and {\bf(B)} NOR gate with the extra balance capacitors. Balance capacitors helps the models to have stable transient current, thus eliminating the metastability during the transitions.}
\label{fig:ratchet_nand_nor}
\end{figure}

In an analogous process, ratchet-based NAND gates and NOR gates can be created with two p-type and two n-type ratchets, with an extra balance capacitor between ratchets in pull-down network of NAND or pull-up network of NOR, to balance the current (as shown in the Fig. \ref{fig:ratchet_nand_nor}). The balance capacitor helps the model to have a stable transient current, so that gates do not have meta-stable states between transitions.

In order to further capture the cascading behavior of the ratchet in our circuit model, we have modified the Schottky diode in our circuit to be {\it{a three terminal device}}, i.e., a voltage controlled rectifier. The third terminal is dependent on the open circuit voltage built across the previous ratchet. This voltage adjusts the turn-on voltage for the device, thus mimicking the on and off states of a ratchet based on the back gate voltage. In Fig. \ref{fig:cascade_ratchet}, we show how the voltage across the second ratchet diode is created by the open circuit voltage from the first ratchet.

\begin{figure} [t]
\centering
\includegraphics[scale=0.24]{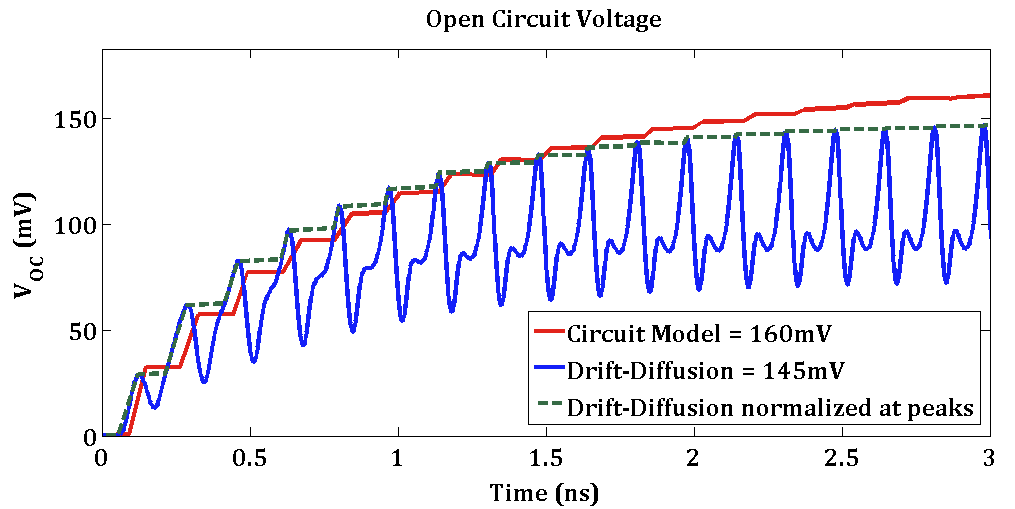}
\caption{Circuit simulation using the circuit diagram in Fig. \ref{fig:circuit} results showing: The voltage being stored on the back gate capacitor compared with the drift-diffusion simulations. For the values of the circuit elements, we used the parameters listed on Table \ref{table2} and for the Schottky parameters, we used barrier height of 15kT and $I_0$ of 1E-15A. Both of the results converge close to $\approx145mV$ and $\approx160mV$, slightly over shooting the expected result of $\approx125mV$. The difference between the curves is our circuit model does not capture the back-flow of the electrons as open-circuit builds-up, so it slightly over shoots the actual voltage built-up.} 
\label{fig:ratchet_sim}
\end{figure}

We simulated the circuit in Fig. \ref{fig:circuit} using Virtuoso Spectre. For the values of the circuit elements, we used the parameters listed on Table \ref{table2}. In order to adjust the clock frequency, we changed the inductance and also the initial current so that the energy stored on the inductor remains constant through all clocking frequencies. We show the charging of the back gate capacitor to the open circuit voltage in Fig. \ref{fig:ratchet_sim}. As predicted from eqn. (\ref{eqn:inequality}), the clock charges the back gate only during the times when the the current is positive (potential is up) and discharges when the current is negative (potential is down). We observe a qualitatively matching $V_{OC}$ built-up between the circuit simulations and the drift-diffusion simulations, although the circuit model can not capture the back-flow of particles when the potential is down. The $V_{OC}$ estimated is slightly higher than the calculated value given the simplicity of circuit model and the ideal diode missing the back-flow shown in Fig. \ref{fig:capcharge}A. 

Fig. \ref{fig:energy} shows the energy dissipated in the resistor as a function of the top gate oscillation frequency. We can see that the energy continues to decrease as we scale down the oscillation frequency as predicted by our adiabatic limit. For high frequencies ($\omega_L \gg \omega_2$), we begin to recover eqn. (\ref{eqn:ER}) of the conventional charging mechanism.  

\section{Limitations and Challenges}
We saw how electronic ratchets can be used to perform logic operations {\it{in principle}} at low energy costs, and how energy dissipation can be further reduced by the adiabatic charging process in the absence of timing constraints. However, there are a few critical effects that can still limit the overall performance of the proposed ratchet. Innovative material and device engineering maybe able to circumvent these limitations.

In Section IV, we demonstrated that the energy dissipated by the ratchet can be arbitrarily reduced by turning down the clocking frequency; however there is a lower limit to the frequency  in order to have the ratchet operational. It is worth remembering that the ratchet is ultimately a {\it{non-equilibrium}} device. If we allow the electrons to continuously diffuse by lowering the asymmetric saw-tooth potential too long, we will reach a near homogeneous distribution for which the drift current component upon raising the barriers will be much smaller. This can be seen in eqn. \ref{eqn:current}: as we increase $t_{off}$, both error function terms become small, and their difference decreases. Since electrons move at high speeds through a small-channel high-mobility material, this requires a minimum frequency fast enough before the electron distribution becomes homogenized. In other words, we have to maintain a balance between the mobility of the channel material and the clocking frequency in order to get optimal performance from the ratchet.

As we scale ratchets towards nano-scale dimensions, we invariably reduce the distance between barriers. The clocking frequencies must be increased significantly to maintain the ratchet operational at small dimensions. For a nanometer-sized device, our calculations and simulations predict that the frequencies must be in the Terahertz range in order to generate enough current to drive the next gate. This can be bypassed by slowing down the electrons using a multi-phased clock. A peristaltic ratchet uses multiple clocks with different spatial and temporal phases to achieve directionality. Initially, the electrons are localized by the top gate clock and when the barrier is lowered, they start to diffuse through the channel. At this point, a  second clock raises another set of barriers, slightly offset spatially, that begin to localize the electrons once again. Because of the spatial offset and temporal phases, the electrons have a net directionality at lower clock frequencies.

Aside from the issue of compatibility between intrinsic and imposed frequencies, there is also a design issue that ratchets need to overcome. While the creation of a clock-driven current source in the absence of a drain bias seems like a distinct net positive, by transducing that current into a capacitive voltage (influenced by a design that is ultimately CMOS-centric), we eventually tend to neutralize that advantage. As the ratchet system pumps electrons into the load capacitor, it builds up an open circuit voltage over time. This built-up voltage, however, acts as an effective drain bias which reintroduces the static dissipation into the logic circuit. Therefore, an alternative method of implementing logic circuits should be devised to mitigate this issue. It might be possible to create a current-based logic system which would eschew the static dissipation, at least for several intermediate logic stages. For now, we present this simple method as a demonstration of ratchet logic design, but ref. \cite{saripalli} shows that power efficient computation with limited number of electrons can be achieved in a ratchet-like geometry through the use of purely current driven Binary Decision Design (BDD) Circuits. 

\section{Conclusions}
Biological systems are rife with examples where ratchet-like mechanisms are purported to shuttle elements (typically motor proteins) in the absence of global potentials, powered by non-equilibrium signals such as the hydrolysis of ATP over a finite noise frequency range \cite{Oster1999R67}. The purpose of this paper was four-fold: (a) to extend this concept to a solid-state device, using a clock and a series of potentials created by interdigitated electrodes; (b) to show how such a ratchet device can be cascaded to create universal Boolean logic; (c) to explain the possible energy-saving advantages raising from the reduced dynamic and static dissipation; and (d) finally, to outline limitations exposed by the simulations of our studied device geometries, namely, the importance of switching in future to current-driven (as opposed to voltage-driven) logic, and the need to achieve a temporal matching or resonance between driving and driven frequencies.  

\section{Acknowledgements}
This material is based on work supported by the Nanoelectronics Research Initiative (INDEX Center), ViNC and NanoSTAR. We would also like to acknowledge Suman Datta and Arun Thathachary for useful discussions.  

\bibliographystyle{ieeetr}

\begin{thebibliography}{10}

\bibitem{zhirnov}
V.~Zhirnov, I.~Cavin, R.K., J.~Hutchby, and G.~Bourianoff, ``Limits to binary
  logic switch scaling - a gedanken model,'' {\em Proceedings of the IEEE},
  vol.~91, pp.~1934--1939, nov 2003.

\bibitem{itrs}
``The international technology roadmap for semiconductors report 2010 update
  overview,'' 2010.

\bibitem{feynman}
R.~Feynman, R.~Leighton, and M.~Sands, {\em The Feynman Lectures on Physics},
  vol.~1.
\newblock Boston: Addison-Wesley, second~ed., 1963.

\bibitem{magnasco}
M.~O. Magnasco, ``Forced thermal ratchets,'' {\em Phys. Rev. Lett.}, vol.~71,
  pp.~1477--1481, Sep 1993.

\bibitem{astumian}
R.~D. Astumian, ``Thermodynamics and kinetics of a brownian motor,'' {\em
  Science}, vol.~276, no.~5314, pp.~917--922, 1997.

\bibitem{rice}
S.~Rice, A.~W. Lin, D.~Safer, C.~L. Hart, N.~Naber, B.~O. Carragher, S.~M.
  Cain, E.~Pechatnikova, E.~M. Wilson-Kubalek, M.~Whittaker, E.~Pate, R.~Cooke,
  E.~W. Taylor, R.~A. Milligan, and R.~D. Vale, ``A structural change in the
  kinesin motor protein that drives motility,'' {\em Nature}, vol.~402,
  no.~6763, pp.~778--784, 1999.

\bibitem{dekker}
C.~Dekker, ``Solid-state nanopores,'' {\em Nat Nano}, vol.~2, no.~4,
  pp.~209--215, 2007.

\bibitem{jung}
P.~Jung, J.~G. Kissner, and P.~Hanggi, ``Regular and chaotic transport in
  asymmetric periodic potentials: Inertia ratchets,'' {\em Phys. Rev. Lett.},
  vol.~76, pp.~3436--3439, Apr 1996.

\bibitem{borromeo}
M.~Borromeo and F.~Marchesoni, ``Asymmetric confinement in a noisy bistable
  device,'' {\em Europhysics Letters}, vol.~68, no.~6, p.~783, 2004.

\bibitem{reguera}
D.~Reguera, G.~Schmid, P.~S. Burada, J.~M. Rubi, P.~Reimann, and P.~Hanggi,
  ``Entropic transport: Kinetics, scaling, and control mechanisms,'' {\em Phys.
  Rev. Lett.}, vol.~96, p.~130603, Apr 2006.

\bibitem{pierret}
R.~Pierret, {\em Semiconductor Device Fundamentals}.
\newblock Pearson Education, 1996.

\bibitem{dattabook}
S.~Datta, {\em Electronic Transport in Mesoscopic Systems}.
\newblock Cambridge Studies in Semiconductor Physics and Microelectronic
  Engineering, Cambridge University Press, 1997.

\bibitem{knorren}
R.~Knorren and K.~H. Bennemann, ``Dynamics of excited electrons in copper and
  ferromagnetic transition metals: Theory and experiment,'' {\em Phys. Rev.
  B.}, vol.~61, pp.~9427--9440, Apr 2000.

\bibitem{yeo}
K.-S. Yeo and K.~Roy, {\em Low Voltage, Low Power VLSI Subsystems}.
\newblock New York, NY, USA: McGraw-Hill, Inc., 1~ed., 2005.

\bibitem{tnano11}
M.~Kabir, D.~Unluer, L.~Li, A.~Ghosh, and M.~Stan, ``Electronic ratchet: A
  non-equilibrium, low power switch,'' in {\em Nanotechnology (IEEE-NANO), 2011
  11th IEEE Conference on}, pp.~482 --486, aug. 2011.

\bibitem{denker}
J.~Denker, ``A review of adiabatic computing,'' in {\em Low Power Electronics,
  1994. Digest of Technical Papers., IEEE Symposium}, pp.~94--97, oct 1994.

\bibitem{takashi}
N.~Anuar, Y.~Takahashi, and T.~Sekine, ``Two phase clocked adiabatic static
  cmos logic,'' in {\em System-on-Chip, 2009. SOC 2009. International Symposium
  on}, pp.~083--086, oct. 2009.

\bibitem{blanchard1941history}
J.~Blanchard, {\em The History of Electrical Resonance}.
\newblock Bell telephone laboratories Incorporated, 1941.

\bibitem{saripalli}
V.~Saripalli, V.~Narayanan, and S.~Datta, ``Ultra low energy binary decision
  diagram circuits using few electron transistors,'' in {\em Nano-Net}, vol.~20
  of {\em Lecture Notes of the Institute for Computer Sciences, Social
  Informatics and Telecommunications Engineering}, pp.~200--209, Springer
  Berlin Heidelberg, 2009.
\newblock 10.1007/978-3-642-04850-0-27.

\bibitem{Oster1999R67}
G.~Oster and H.~Wang, ``Atp synthase: two motors, two fuels,'' {\em Structure},
  vol.~7, no.~4, pp.~R67 -- R72, 1999.

\end{thebibliography}

\end{document}